\documentclass[a4paper]{cas-sc}

\usepackage[authoryear]{natbib}
\usepackage{float}
\def\tsc#1{\csdef{#1}{\textsc{\lowercase{#1}}\xspace}}
\tsc{WGM}
\tsc{QE}
\usepackage{caption}
\usepackage{graphicx}
\usepackage{booktabs}
\usepackage{multirow}
\usepackage{placeins}
\newenvironment{inlinefigure}{\begin{figure}[pos=!htbp]\centering}{\end{figure}}
\renewcommand{\topfraction}{0.95}
\renewcommand{\bottomfraction}{0.85}
\renewcommand{\textfraction}{0.05}
\renewcommand{\floatpagefraction}{0.80}

\ExplSyntaxOn

\ExplSyntaxOff

\begin{document}
\let\WriteBookmarks\relax
\renewcommand{\topfraction}{0.95}
\renewcommand{\bottomfraction}{0.90}
\renewcommand{\textfraction}{0.05}
\renewcommand{\floatpagefraction}{0.80}
\renewcommand{\dbltopfraction}{0.95}
\renewcommand{\dblfloatpagefraction}{0.80}
\setcounter{topnumber}{4}
\setcounter{bottomnumber}{3}
\setcounter{totalnumber}{6}
\setcounter{dbltopnumber}{2}
\setlength{\textfloatsep}{8pt plus 2pt minus 2pt}
\setlength{\floatsep}{8pt plus 2pt minus 2pt}
\setlength{\intextsep}{8pt plus 2pt minus 2pt}
\makeatletter
\setlength{\@fptop}{0pt}
\setlength{\@fpsep}{8pt}
\setlength{\@fpbot}{0pt plus 1fil}
\makeatother

\shorttitle{}    

\shortauthors{X. Liu et al.}  

\title [mode = title]{Balancing Mobility Efficiency and Environmental Exposure During Mega Events: A Multi-agent Routing Framework}  



\author[1]{Xinyue Liu}
\author[1,2]{Qiumeng Li}
\cormark[1]
\author[3]{Suhong Zhou}
\author[4]{Junwen Lu}
\author[5]{Jiali Zhou}
\author[3]{Weipan Xu}

\affiliation[1]{organization={Thrust of Urban Governance and Design, Society Hub, The Hong Kong University of Science and Technology (Guangzhou)},
                city={Guangzhou},
                postcode={511453},
                country={China}}
\affiliation[2]{organization={Thrust of Intelligent Transportation, Systems Hub, The Hong Kong University of Science and Technology (Guangzhou)},
                city={Guangzhou},
                postcode={511453},
                country={China}}
\affiliation[3]{organization={School of Geography and Planning, Sun Yat-sen University},
                city={Guangzhou},
                postcode={510006},
                country={China}}
\affiliation[4]{organization={School of Intelligent Systems Engineering, Sun Yat-sen University},
                city={Guangzhou},
                postcode={510006},
                country={China}}
\affiliation[5]{organization={Department of Urban Planning and Design, The University of Hong Kong},
                addressline={8/F Knowles Building, Pokfulam Road},
                city={Hong Kong},
                citysep={}}

\cortext[1]{Corresponding author: Qiumeng Li. Email: \href{mailto:qiumengli@hkust-gz.edu.cn}{qiumengli@hkust-gz.edu.cn}.}

\begin{abstract}
Mega events can generate concentrated travel demand, intensifying congestion around venues and increasing travelers’ environmental exposure. Existing traffic management strategies focus on travel efficiency, with limited attention to exposure risks. Taking mega-event travel as a short-term high-demand scenario, this study develops an exposure-aware multi-agent routing framework integrating road-network conditions, event-induced demand, background traffic, and environmental data within a unified simulation environment. Motorized travelers optimize travel efficiency, whereas micro-mobility travelers consider travel time and dynamic exposure to traffic-related PM2.5 and noise. We compare the framework with shortest-distance, congestion-aware, and exposure-aware routing baselines. Results show that, under high demand, the SARSA-based framework reduces vehicle travel time by 34.59\% and cumulative environmental exposure by 58.65\%, while micro-mobility travel time remains nearly unchanged. Exposure reduction is jointly driven by vehicle-side congestion mitigation and low-exposure routing. The framework provides quantitative support for vehicle diversion, low-exposure route guidance, and health-oriented traffic management in short-term high-demand scenarios.
\end{abstract}




\maketitle
\section{Introduction}

Urban transportation has important environmental and health impacts, particularly through air pollution and traffic noise \citep{askariyeh2020near, levy2010evaluation}. In the context of sustainable transportation, urban transport systems are increasingly expected to go beyond improving mobility efficiency and reducing travel time to also consider environmental quality. This perspective is consistent with the United Nations Sustainable Development Goals, particularly SDG 11.2, which emphasizes safe, accessible, and sustainable transport systems, and SDG 3.9, which highlights the need to reduce health burdens associated with environmental pollution. Traffic-related PM$_{2.5}$ and noise exposure are closely associated with traffic volume, vehicle speed, and congestion \citep{hilker2019traffic, seto2007spatial}. When large numbers of people travel and gather within limited spatial and temporal ranges, rapidly increasing travel demand can aggravate congestion and pollutant emissions while exposing more travelers to polluted and noisy environments. Therefore, reducing environmental exposure while maintaining transportation efficiency is an important issue for health-oriented and sustainable transport management.

Mega events, such as sports games and concerts, represent typical scenarios characterized by intensive crowd gathering and highly concentrated travel demand. Because large numbers of visitors usually travel to a limited number of venues within narrow time windows, mega-event travel is often characterized by highly synchronized demand, spatially concentrated destinations, and rapid increases in traffic volume. Previous studies have shown that mega events can substantially influence urban PM$_{2.5}$ concentrations and local air quality, while event-related traffic may generate short-term pollution hotspots around venues and along major travel corridors \citep{li2021megaevents, peel2010impact}. Meanwhile, concentrated increases in vehicle traffic may also increase road traffic noise and travelers' environmental exposure along their routes. Compared with regular peak periods, traffic demand during mega events is generally more abrupt and spatially concentrated, often resulting in severe congestion, longer and more uncertain travel times, and increased traffic-related environmental exposure risks \citep{villiers2019traffic,xu2017collective}. Therefore, mega events can be regarded as representative concentrated-demand scenarios in which transportation efficiency and environmental exposure are simultaneously placed under greater pressure, providing a typical case for examining their joint optimization.

Despite growing attention to these issues, most existing route optimization and traffic management studies still primarily focus on minimizing travel time and travel cost, with relatively limited consideration of environmental health. Although exposure-aware routing has received increasing attention, existing approaches often rely on static or externally specified exposure information, with limited representation of dynamically evolving exposure conditions shaped by route choice and traffic states. In addition, existing routing studies often focus on a single travel mode, despite substantial differences in travel costs and environmental exposure characteristics between vehicle travelers and micro-mobility users. Accordingly, this study addresses one central research question: Under concentrated travel demand, how can environmental exposure be incorporated into multi-agent route planning to coordinate transportation efficiency across travel modes while reducing exposure for micro-mobility travelers, thereby enabling joint improvements in efficiency and exposure? 

Against this background, this study develops an exposure-aware multi-agent route planning framework with differentiated objectives for vehicle and micro-mobility agents. The framework considers transportation efficiency for both modes and environmental exposure for micro-mobility users, enabling the joint optimization of transportation efficiency and environmental health.

The main contributions of this study are threefold. First, we incorporate environmental exposure into multi-agent route optimization for short-term, high-demand travel scenarios. This extends conventional efficiency-oriented routing toward the joint consideration of transportation efficiency and environmental health. Second, we introduce a dynamic traffic–exposure feedback mechanism. Route choices reshape traffic conditions, which then affect subsequent travel costs and environmental exposure. This allows route choice, traffic states, and exposure to co-evolve during the simulation. Third, the proposed framework provides a practical basis for health-oriented transport applications. It can support low-exposure navigation, greenway planning, and walking and cycling network design. 

\section{Related Work}

Traffic route optimization is a fundamental topic in transportation planning and management. Traditional route-planning and traffic-assignment methods typically aim to minimize travel distance, travel time, or generalized travel cost through deterministic shortest-path algorithms, user-equilibrium models, and system-optimal assignment formulations \citep{wardrop1952theoretical,sheffi1985urban,patriksson1994assignment}. These methods provide the theoretical foundation for modeling route choice and network loading, but they often assume stable equilibrium conditions, complete information, or centralized optimization. In practice, travelers make autonomous route choices under incomplete information, and their decisions interact through congestion and network-capacity constraints \citep{grunitzki2016combining,shou2022multi}. Recent studies have increasingly applied multi-agent reinforcement learning to adaptive traffic assignment and route choice under large and dynamically varying demand \citep{wang2025scalable,cun2026multimodal}. In particular, mean-field formulations have been used to simplify large-scale agent interactions and support multi-modal traffic assignment with heterogeneous route and mode choices \citep{cun2026multimodal}. Consequently, route optimization is increasingly moving toward adaptive decision frameworks that explicitly represent interactions among travelers and evolving network states.

Growing concern about sustainable transportation and public health has motivated increasing research on environmentally sensitive and exposure-aware route planning. Previous studies have shown that route choice can substantially affect travelers' exposure to traffic-related air pollution, particularly for pedestrians and cyclists who are directly exposed to roadside environments \citep{hertel2008human,hatzopoulou2013web,hofman2018cyclist,rafiepourgatabi2021route}. Low-exposure routes can reduce exposure to pollutants such as black carbon, ultrafine particles, NO$_2$, and PM$_{2.5}$ compared with traffic-intensive, shortest, or fastest routes \citep{hertel2008human,hofman2018cyclist,wu2022pm25}. Recent work has further integrated high-resolution environmental prediction with route optimization. For example, Geo-AI-based street-level NO$_2$ modeling has been combined with exposure-aware routing to identify lower-exposure commuting alternatives at fine spatial resolution \citep{lee2026exposure}. Beyond air pollution, recent studies have incorporated multiple environmental factors, including traffic noise, street-level greenery, and other characteristics of active-travel environments, into route assessment and optimization \citep{willberg2023cyclists,helle2023green}. These approaches generally extend shortest-path or least-cost-path algorithms by incorporating pollutant concentrations, noise levels, greenery, or environmental quality into link-level impedance, thereby identifying routes that reduce cumulative exposure with relatively modest increases in travel distance or travel time \citep{hatzopoulou2013web,helle2023green,lee2026exposure}.

Despite these advances, most exposure-aware routing studies still focus on individual route choice, active travel, or relatively static environmental cost surfaces. Environmental exposure is commonly represented as link-level impedance derived from pre-estimated air-pollution, noise, greenery, or other environmental data \citep{hatzopoulou2013web,alam2018comparison,rafiepourgatabi2021route,helle2023green}. Under this approach, exposure is mainly treated as a spatial attribute attached to road links, while the feedback among route choice, congestion evolution, and changing exposure conditions is generally not explicitly considered. However, dynamic traffic-assignment studies have shown that route choices and network traffic states are mutually dependent, and that congestion can influence both travel time and emission-related exposure conditions \citep{peeta1995system,wang2018dynamic,zhang2013air}. More recent MARL-based studies further demonstrate that large-scale and multi-modal routing can be formulated as interacting sequential decisions in which agent choices continuously reshape network states \citep{wang2025scalable,cun2026multimodal}. When many travelers respond to the same routing objective, their collective choices may transfer congestion to alternative links, alter local traffic density, and redistribute environmental exposure risks. Exposure-aware routing under high-demand conditions therefore requires a dynamic framework that captures the interactions among travelers, evolving traffic states, and environmental costs.

In mega-event scenarios, existing transportation studies have mainly focused on improving travel efficiency through traffic-demand management, traveler-information provision, temporary traffic control, public-transport coordination, and network-operation strategies \citep{giuliano1990staggered,xing2025empirical}. Recent event-aware mobility research further shows that concerts and other public events can generate irregular and highly concentrated travel-demand surges, requiring coordinated transport responses across modes and agencies \citep{wang2026eventaware}. Mega events can generate highly concentrated travel demand within short time periods and around specific venues, placing substantial pressure on road networks and public-transport systems. Although such events may also produce short-term changes in traffic-related air pollution and environmental noise, exposure-related costs are rarely incorporated directly into event-related routing or traffic-management strategies. More importantly, most routing studies treat travelers as a homogeneous group and pay limited attention to the different travel costs experienced by vehicle users and micro-mobility users. Vehicle users are primarily affected by congestion-induced delays, whereas micro-mobility users are directly exposed to roadside air pollution and traffic noise during travel. Existing studies also seldom integrate these differentiated travel costs with the dynamic feedback among route choice, congestion evolution, and environmental exposure within a unified spatial simulation framework.

\section{Method}
\label{sec:revised_method}

\subsection{Research Framework}

This study proposes an exposure-aware multi-agent routing framework for mega-event traffic management. As shown in Figure~\ref{fig:research_framework}, real-world traffic and environmental data are integrated into a unified grid-based environment. Vehicle agents optimize travel time, while micro-mobility agents consider both travel time and environmental exposure. The framework is evaluated under different demand and exposure settings in terms of traffic efficiency and exposure reduction.

\begin{inlinefigure}
\centering
\includegraphics[width=\linewidth]{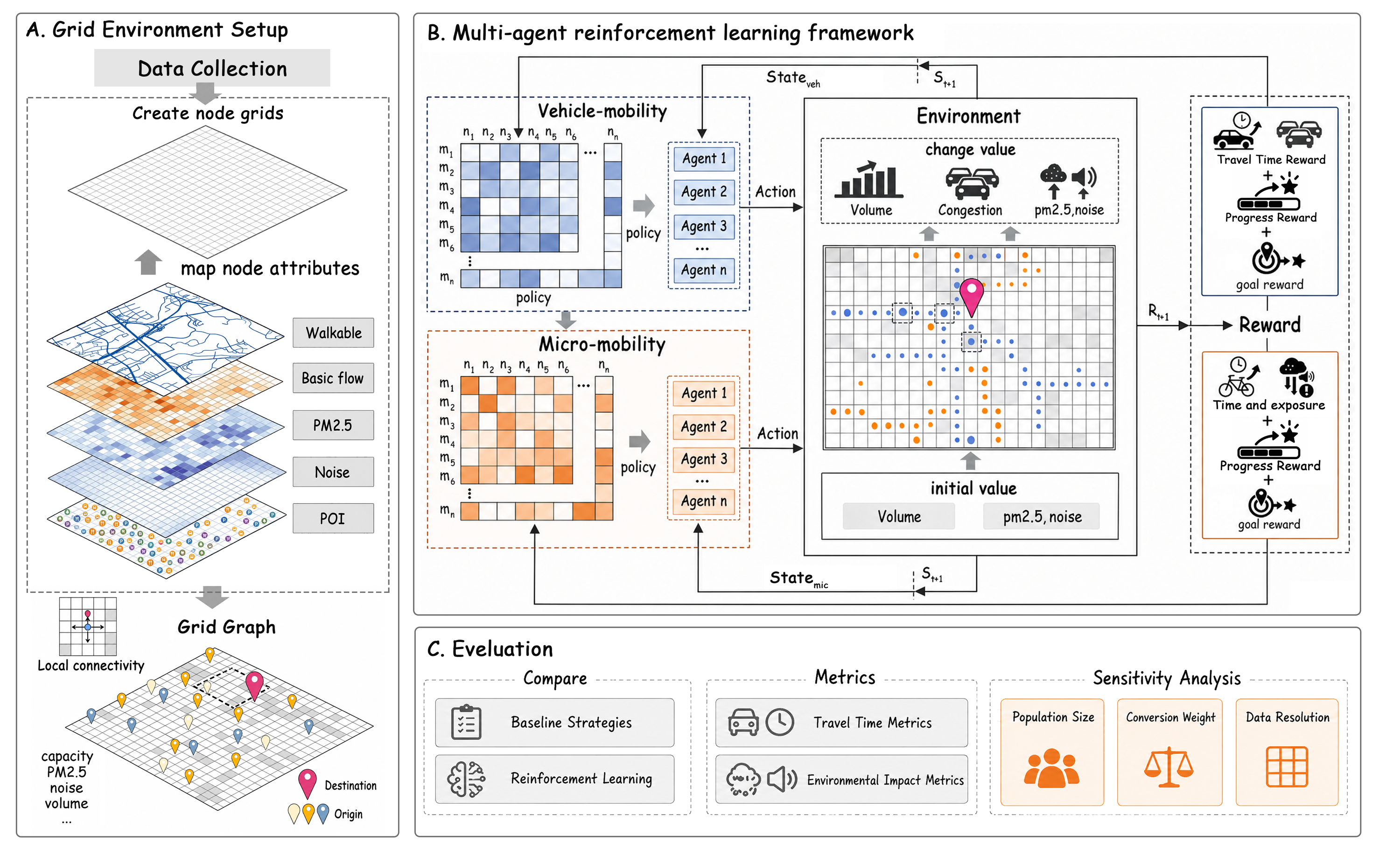}
\caption{Research framework.}
\label{fig:research_framework}
\end{inlinefigure}

\subsection{Study Area and Data}

This study focuses on the Olympic Sports Center and its surrounding road network as the study area. As a major venue for sports games, concerts, and other large-scale public events, this area often experiences short-term surges in travel demand during event periods. Such concentrated and time-sensitive demand places substantial pressure on the surrounding road network, making it a representative scenario for analyzing event-induced congestion and environmental exposure. The simulation is scenario-based and represents event-induced travel demand rather than a direct reconstruction of a specific real event.

To construct the simulation environment, this study integrates four types of data. Road network data from OpenStreetMap are used to construct the traversable road grid and estimate road-segment attributes, including link length, connectivity, and capacity. Point-of-interest (POI) data from Gaode Map, including residential areas, hotels, and transport stations, are used to identify potential internal origins. Mobile signaling records collected on a regular weekday from 7:00 to 18:00 are aggregated into hourly movement counts to represent the background traffic demand under normal conditions, which serves as the baseline traffic flow in the simulation. The environmental inputs are derived from monitoring-station data and provided as hourly citywide gridded products for Guangzhou, including PM$_{2.5}$ concentration in $\mu$g\,m$^{-3}$ and equivalent continuous sound level (LEQ) in dB.

\subsection{Simulation Environment}

The real urban road network is mapped onto a regular grid with equal-sized cells. Only cells containing road segments are retained as traversable units. Each traversable cell is assigned traffic and environmental attributes, including road connectivity, capacity, background traffic conditions, PM$_{2.5}$ concentration, and noise level. Road-network data, mobile signaling data, POIs, and environmental data are spatially matched within the same grid system. This allows traffic conditions, environmental attributes, and agent locations to be updated consistently during the simulation. The primary simulation uses a 250 m grid. To examine the robustness of the framework to spatial resolution, additional experiments are conducted using 100, 150, and 200 m grids, with the same spatial construction and simulation rules applied across resolutions.

The Olympic Sports Center is defined as the common destination for all event-related trips. Trip origins are generated through spatially weighted sampling. The sampling considers both population activity intensity derived from mobile signaling data and the spatial distribution of POIs. This allows the simulated demand to reflect spatial differences in human activity and potential trip generation across the study area.

Demand $N$ represents the additional event-related travel demand superimposed on the baseline urban traffic represented by the mobile signaling data. The Olympic Sports Center has a maximum capacity of approximately 80,000 people. This venue capacity is used to define the upper bound of the event-demand scenarios, assuming that event attendees generate additional trips toward the venue during the modeled pre-event period. The motorized travel share is set to 23.8\%, following the 2023 Guangzhou Transport Development Annual Report, which reports that individual motorized travel accounted for 23.8\% of trips in central Guangzhou \citep{guangzhou2024transport}. Reported passenger-car occupancies vary with travel context but commonly fall between one and two persons per vehicle. Values of approximately 1.4--2.0 have been reported for different trip purposes in Chinese cities. Accordingly, the mean vehicle occupancy is set to 1.5 persons per vehicle. Motorized demand is first determined by the travel-mode share and then converted into the number of vehicle agents based on the average vehicle occupancy. The remaining demand is assigned to micro-mobility. One vehicle agent represents one vehicle, while one micro-mobility agent represents one traveler.

Both travel modes are simulated on the same physical clock with a time step of $\Delta t=2.5$ s. To represent gradual arrivals before the event, each agent is assigned a departure time within the first 1,800 s and enters the system accordingly. Each agent can remain in the simulation for up to 7,200 s after departure, including waiting and congestion-related delays, and the simulation ends when all agents either reach the Olympic Sports Center or reach their maximum simulation duration. Free-flow speeds are set to 40 km\,h$^{-1}$ for vehicles and 10 km\,h$^{-1}$ for micro-mobility users. At each routing decision, an agent selects an adjacent traversable cell and moves toward it according to its current speed. If a downstream cell reaches its capacity, a vehicle waits in the current cell. Routing decisions are updated only after the current cell-to-cell movement is completed and the agent can proceed.

\subsection{Traffic-Related Environmental Exposure Modeling}

The environmental module converts evolving traffic states into event-related PM$_{2.5}$ and traffic-noise increments on the common simulation clock. It is designed to provide a consistent reduced-form environmental field for comparing routing strategies under the same observed background conditions, rather than to reproduce absolute short-term pollutant concentrations or site-specific acoustic propagation.

Let $C(g)$ denote the simultaneous vehicle storage of grid cell $g$, $b(g)$ its background vehicle occupancy, and $n_t(g)$ the number of admitted event vehicles. The occupancy ratio and corresponding vehicle speed are
\begin{equation}
\rho_t(g)=\frac{b(g)+n_t(g)}{C(g)},\qquad
v_t(g)=\frac{v_0}{1+0.15\rho_t(g)^4},
\label{eq:congestion_ratio}
\end{equation}
where $v_0=40$ km h$^{-1}$. Following the conventional Bureau of Public Roads (BPR) volume--delay formulation \citep{bpr1964}, this BPR-shaped relationship is used as a grid-level delay proxy because $C(g)$ represents spatial storage rather than hourly flow capacity.

Moving vehicles contribute speed-dependent PM$_{2.5}$ emissions and rolling-plus-propulsion noise, whereas vehicles blocked within the road network contribute idling PM$_{2.5}$ emissions and propulsion noise. For a moving gasoline passenger car, the PM$_{2.5}$ emission rate is
\begin{equation}
q^P(v)=e_0\phi\frac{v/3.6}{1000},\qquad
e_0=3.0\ \mathrm{mg\,vehicle^{-1}\,km^{-1}},\quad \phi=0.32,
\label{eq:pm_moving_rate}
\end{equation}
where $v$ is expressed in km h$^{-1}$. The correction factor $\phi=0.32$ corresponds to the reference speed of 40 km h$^{-1}$ \citep{mee2014roadvehicle}. A stationary on-road vehicle contributes $q^P_{\mathrm{idle}}=0.0248$ mg s$^{-1}$ \citep{park2011high}.

At the grid level, moving and idling emissions are normalized by a common traffic reference:
\begin{equation}
u^P_t(g)=
\frac{
\displaystyle\sum_{i\in\mathcal V^{\mathrm{mov}}_t(g)}q^P(v_{i,t})+
\displaystyle\sum_{i\in\mathcal V^{\mathrm{idle}}_t(g)}q^P_{\mathrm{idle}}
}{C(g)q^P(v_0)}.
\label{eq:pm_source_ratio}
\end{equation}
Here, $C(g)q^P(v_0)$ represents the emission rate when the cell reaches its vehicle storage capacity and all vehicles travel at $v_0$. Thus, $u^P_t(g)$ represents capacity-normalized traffic-emission intensity.

The normalized intensity is mapped onto the observed hourly background concentration as
\begin{equation}
P_t(g)=P_0(g)\left[1+\kappa_Pu^P_t(g)\right],
\qquad \kappa_P=0.131,
\label{eq:pm_background_anchor}
\end{equation}
where $P_0(g)$ is the observed hourly background PM$_{2.5}$ concentration. The parameter $\kappa_P$ is anchored to the 2023 Guangzhou source-apportionment estimate that attributed 13.1\% of ambient PM$_{2.5}$ to motor vehicles \citep{guangzhou2025pm25}. It therefore provides an empirical scaling reference between simulated traffic-emission intensity and event-related PM$_{2.5}$ increments.

Traffic noise is modeled using the CNOSSOS-EU category-1 rolling and propulsion equations \citep{kephalopoulos2012cnossos}. Let $W_A(v)$ denote the A-weighted sound-energy source of a moving vehicle at speed $v$. Because the standard equations do not define a zero-speed road-vehicle source, stationary on-road vehicles are assigned a separate source $W_{A,\mathrm{idle}}$, defined by the propulsion component at the 20 km h$^{-1}$ lower modeling boundary.

Because decibel levels are logarithmic sound-energy ratios, background and event-related contributions are combined on the linear energy scale before conversion back to decibels. The capacity-normalized acoustic loading and resulting noise level are
\begin{equation}
u^L_t(g)=
\frac{
\displaystyle\sum_{i\in\mathcal V^{\mathrm{mov}}_t(g)}W_A(v_{i,t})+
\displaystyle\sum_{i\in\mathcal V^{\mathrm{idle}}_t(g)}W_{A,\mathrm{idle}}
}{C(g)W_A(v_0)},
\qquad
L_t(g)=L_0(g)+10\log_{10}\!\left[1+\kappa_Lu^L_t(g)\right],
\quad \kappa_L=1,
\label{eq:noise_background_anchor}
\end{equation}
where $L_0(g)$ is the observed hourly background noise level. The parameter $\kappa_L=1$ defines the normalized acoustic specification used in the simulation.

The routing signal is based on the additional threshold exceedance associated with event traffic. We use $P_{\mathrm{th}}=15$ $\mu$g\,m$^{-3}$ for PM$_{2.5}$ and $L_{\mathrm{th}}=70$ dB for noise \citep{who2021airquality,gb3096_2008}. The PM$_{2.5}$ threshold follows the WHO 24-h guideline, while the noise threshold follows the corresponding regulatory reference.

With $[x]_+=\max(x,0)$, the event-related incremental burdens are
\begin{equation}
E^{P,\Delta}_t(g)=
\left[[P_t(g)-P_{\mathrm{th}}]_+-[P_0(g)-P_{\mathrm{th}}]_+\right]_+,
\qquad
E^{L,\Delta}_t(g)=
\left[[L_t(g)-L_{\mathrm{th}}]_+-[L_0(g)-L_{\mathrm{th}}]_+\right]_+.
\label{eq:incremental_exposure_components}
\end{equation}

The two components are normalized by their reference levels and combined without additional weighting:
\begin{equation}
I^\Delta_t(g)=
\frac{E^{P,\Delta}_t(g)}{P_{\mathrm{th}}}+
\frac{E^{L,\Delta}_t(g)}{L_{\mathrm{th}}}.
\label{eq:exposure_index}
\end{equation}
The equal numerical weighting places PM$_{2.5}$ and noise on a common scale and does not imply equal health-effect weights. The resulting $I^\Delta_t(g)$ is a dimensionless comparative exposure index representing the instantaneous environmental burden associated with event traffic.

\subsection{Reinforcement-Learning Formulation}

This study uses SARSA and MAPPO to implement traffic--environmental exposure routing optimization and to compare the performance of different policy representations. SARSA performs tabular value learning based on discretized grid states, whereas MAPPO uses neural networks for multi-agent policy learning. Both methods are trained and evaluated within the same traffic and environmental exposure environment.

\paragraph{State.}

To keep the state representation scalable under large agent populations, we use a mean-field representation that summarizes nearby interactions through local aggregate traffic conditions rather than the states of individual agents \citep{yang2018mean}. Each agent state includes its current grid location, a mode-specific local condition, and the mean vehicle occupancy of the surrounding $3\times3$ neighbourhood.

Let $\bar\rho_t(g)$ denote the mean vehicle occupancy ratio over all traversable cells within the $3\times3$ neighbourhood of grid $g$ at time $t$. For vehicle agents, the local condition is the occupancy ratio $\rho_t(g)$ of the current grid, whereas for micro-mobility agents it is the environmental exposure index $I^\Delta_t(g)$. The mode-specific states are

\begin{equation}
x^{\mathrm{veh}}_{a,t}=\bigl(
g^a_t,\rho_t(g^a_t),\bar\rho_t(g^a_t)\bigr),
\qquad
x^{\mathrm{mic}}_{a,t}=\bigl(
g^a_t,I^\Delta_t(g^a_t),\bar\rho_t(g^a_t)\bigr).
\label{eq:shared_rl_state}
\end{equation}

For SARSA, the continuous local and neighbourhood variables are discretized into three levels to form a finite tabular state space:

\begin{equation}
\tilde{s}^{\mathrm{veh}}_{a,t}=(g^a_t,b^c_{a,t},b^m_{a,t}),
\qquad
\tilde{s}^{\mathrm{mic}}_{a,t}=(g^a_t,b^e_{a,t},b^m_{a,t}),
\label{eq:sarsa_state}
\end{equation}

where $b^c_{a,t}$, $b^e_{a,t}$, and $b^m_{a,t}$ denote the discretized local congestion level, local exposure level, and neighbourhood traffic condition, respectively.

MAPPO follows the same state-design principle but uses the continuous representation directly. Its actor receives the agent location, mode-specific local condition, and neighbourhood traffic condition, while the centralized critic additionally uses aggregate network-level information during training. Thus, SARSA and MAPPO differ mainly in policy representation: SARSA uses tabular on-policy value learning over discretized states, whereas MAPPO uses neural-network-based actor--critic learning over continuous states.

\paragraph{Action.}

The basic action space consists of movements to the four orthogonally adjacent traversable cells. Actions leading outside the connected road network are masked. When another legal action is available, an immediate reversal to the previous cell is also excluded to reduce unnecessary backtracking. Vehicles can select among all legal neighbouring cells. For micro-mobility users, the action space is further constrained by route length. A legal action must reduce the remaining shortest-path distance to the destination. Micro-mobility users therefore choose among equal-length or equivalent shortest-route alternatives without increasing the shortest-path distance. This constraint preserves basic travel efficiency while allowing the agent to identify lower-exposure alternatives among feasible routes with comparable route length. It also prevents exposure reductions from being achieved primarily through excessive detours. Each action selects the next adjacent cell and initiates movement toward that cell, rather than producing an instantaneous cell-to-cell displacement within a single simulation step.

\paragraph{State Transition.}

At decision epoch $k$, agent $a$ selects action $a_{a,k}$ and moves toward the selected neighbouring cell. A routing transition ends when the cell crossing, including any capacity-induced waiting, is completed. During this process, traffic conditions and the environmental exposure field are updated at every physical time step $\Delta t$, and waiting vehicles continue to occupy road space and contribute idling emissions. The next state is then constructed from the agent's updated location, local traffic or exposure condition, and neighbourhood mean-field condition, after which the next routing decision is made.

\paragraph{Reward.}

The reward functions reflect the different routing objectives of vehicle and micro-mobility agents. The core objective for vehicles is travel-time reduction, whereas micro-mobility agents additionally account for dynamic environmental exposure. Route-progress, arrival, and timeout terms are included as auxiliary shaping components to stabilize learning and promote trip completion.

For vehicle agents, let $\mathcal T_{a,k}$ denote the physical time steps elapsed during routing transition $k$, $D_{a,k}=|\mathcal T_{a,k}|\Delta t$ its duration, and
$\Delta d_{a,k}=d(g^{\mathrm{start}}_{a,k})-d(g^{\mathrm{end}}_{a,k})$
the reduction in remaining shortest-path distance. The vehicle reward is

\begin{equation}
R_{a,k}=
-w_{\mathrm{veh}}\frac{D_{a,k}}{60}
+w^{\mathrm{veh}}_p\Delta d_{a,k}
+R_{\mathrm{ms}}\Delta b_{a,k}
-R_{\mathrm{np}}\mathbf{1}_{\Delta d_{a,k}\leq 0}
+R^{\mathrm{veh}}_{\mathrm{goal}}\mathbf{1}_{\mathrm{arrive}}
-R_{\mathrm{timeout}}\mathbf{1}_{\mathrm{timeout}}.
\label{eq:vehicle_reward}
\end{equation}

The first term penalizes travel time and therefore represents the main vehicle objective. The remaining terms shape route progress and trip completion: $\Delta d_{a,k}$ rewards movement toward the destination, $\Delta b_{a,k}$ rewards improvement beyond the best previously reached distance, and $R_{\mathrm{np}}$ penalizes non-progress. Arrival and timeout terms encourage successful completion within the simulation horizon.

For micro-mobility agents, environmental exposure is added directly to the transition cost:

\begin{equation}
R_{a,k}=
-\sum_{t\in\mathcal T_{a,k}}
\left(w_{\mathrm{mic}}+\eta I^\Delta_t(g^a_t)\right)\frac{\Delta t}{60}
+w^{\mathrm{mic}}_p\Delta d_{a,k}
+R^{\mathrm{mic}}_{\mathrm{goal}}\mathbf{1}_{\mathrm{arrive}}
-R_{\mathrm{timeout}}\mathbf{1}_{\mathrm{timeout}}.
\label{eq:micromobility_reward}
\end{equation}

Here, $w_{\mathrm{mic}}$ controls travel-time cost, while $\eta$ determines the importance of the dynamic exposure index $I^\Delta_t(g^a_t)$. Micro-mobility agents therefore seek lower-exposure routes while maintaining travel efficiency. The progress, arrival, and timeout terms serve the same auxiliary role as in the vehicle reward.

SARSA and MAPPO share this core time--exposure objective, while algorithm-specific shaping parameters are adjusted for training stability. The SARSA reward parameters are reported in Appendix~\ref{app:reward_parameters}.

\subsection{Baseline Routing Strategies}

Five nonlearning routing strategies are constructed to provide matched comparisons under the same travel demand, simulation clock, road-storage constraints, and dynamic environmental conditions (Table~\ref{tab:baseline_configuration}). These baselines progressively introduce congestion responsiveness and environmental exposure into route decisions, allowing the contribution of each component to be distinguished from that of the learned policy.

Shortest-distance routing serves as the reference strategy and minimizes road-graph distance for both vehicle and micro-mobility agents. Congestion-aware routing allows vehicle agents to repeatedly select routes according to the current occupancy-dependent travel time, while micro-mobility agents continue to follow shortest-distance routes. This baseline isolates the effects of vehicle-side traffic redistribution without introducing an explicit exposure objective for micro-mobility users.

The remaining three strategies retain the same congestion-aware vehicle routing and differ only in the treatment of micro-mobility exposure. Static exposure-aware routing minimizes a fixed cost of $1+2I(g)$ based on the background exposure field. Dynamic exposure-aware routing applies the same exposure weighting but updates the remaining route after each completed cell crossing using the current event-related exposure field $I^\Delta_t(g)$. It therefore provides a direct comparison with a responsive but nonlearning exposure-aware routing rule. Minimum-exposure routing prioritizes the fixed background exposure field, with a small positive step cost added to prevent zero-cost loops. This strategy represents an exposure-priority benchmark and is used to examine the travel-time cost associated with pursuing stronger exposure reduction.

These baselines compare shortest-distance routing, vehicle congestion response, exposure-aware micro-mobility routing, and the efficiency cost of stronger exposure prioritization. Among them, dynamic exposure-aware routing is the closest nonlearning comparator because it also uses dynamic exposure information but updates routes directly based on current costs rather than a learned policy.  All baseline strategies are evaluated under the same simulation environment as the learned policies and follow the same feasible-movement and no-immediate-reversal rules.
\begin{table}[t]
\caption{Nonlearning baseline routing strategies.}
\label{tab:baseline_configuration}
\centering
\footnotesize
\setlength{\tabcolsep}{3.5pt}
\begin{tabular}{p{0.24\columnwidth}p{0.30\columnwidth}p{0.36\columnwidth}}
\toprule
Method & Vehicle routing & Micro-mobility routing \\
\midrule

Shortest-distance
& Shortest road-graph distance
& \multirow{2}{=}{Shortest road-graph distance} \\

Congestion-aware
& \multirow{4}{=}{Current congestion-dependent travel time}
& \\

Static exposure-aware
&
& Fixed background exposure-weighted cost \\

Dynamic exposure-aware
&
& Current exposure-weighted cost; replanned after each crossing \\

Minimum-exposure
&
& Fixed background exposure-priority cost \\

\bottomrule
\end{tabular}
\end{table}

\subsection{Evaluation Metrics}

The three primary evaluation metrics are total in-network vehicle time, total micro-mobility travel time, and cumulative incremental exposure.

Let $\mathcal T^{\mathrm{net}}_a$ denote the physical time steps from vehicle admission until the trip ends or reaches the simulation limit. Total in-network vehicle time is

\begin{equation}
T_{\mathrm{veh}}=
\sum_{a\in\mathcal A_{\mathrm{veh}}}
\sum_{t\in\mathcal T^{\mathrm{net}}_a}
\frac{\Delta t}{60}.
\label{eq:vehicle_total_time}
\end{equation}

This metric includes both movement and downstream-blocked waiting within the modeled network. Waiting before network admission is excluded and retained separately as entry-queue time.

Total micro-mobility travel time is

\begin{equation}
T_{\mathrm{mic}}=
\sum_{a\in\mathcal A_{\mathrm{mic}}}
\sum_{t\in\mathcal T_a}
\frac{\Delta t}{60},
\label{eq:micro_total_time}
\end{equation}

where $\mathcal T_a$ denotes the active travel period of micro-mobility user $a$.

Cumulative incremental exposure is obtained by integrating the dynamic exposure index along each micro-mobility trajectory:

\begin{equation}
EI_a^\Delta=
\sum_{t\in\mathcal T_a}
I^\Delta_t(g^a_t)\frac{\Delta t}{60},
\qquad
EI^\Delta=
\sum_{a\in\mathcal A_{\mathrm{mic}}}EI_a^\Delta.
\label{eq:system_incremental_exposure}
\end{equation}

The population total $EI^\Delta$ is reported in index-minutes and is interpreted as a comparative measure of event-related environmental burden rather than an individual health dose. PM$_{2.5}$ concentration--time exposure, equivalent noise exposure, entry-queue time, and in-network blocked time are retained as diagnostic indicators.

Percentage changes are calculated relative to the matched shortest-distance baseline under the same demand level and grid resolution, with positive values indicating reductions. Percentage changes are not reported when the reference value is zero.

\subsection{Experimental Settings}

All reinforcement-learning experiments use five training seeds. In the main experiment, SARSA and MAPPO are trained for 100 episodes. For SARSA, the learning rate and exploration rate decay during the early stage of training, while MAPPO is optimized using Adam with PPO updates. After training, the selected policies are frozen during evaluation.

The main experiment uses a 250 m grid and five event-demand levels:
$N=40{,}000$, $50{,}000$, $60{,}000$, $70{,}000$, and $80{,}000$.
Deterministic routing baselines, SARSA, and MAPPO are compared at each demand level, with the main exposure weight set to $\eta=3$.

At $N=80{,}000$, sensitivity analyses examine the exposure weight $\eta\in\{0,1,3,5,7\}$ and route-guidance obedience from 0 to 100\% in 25-percentage-point increments. Additional experiments at 100, 150, and 200 m are included to examine sensitivity to spatial resolution. The learning model is retrained for each exposure-weight setting, whereas the obedience analysis keeps the trained main-experiment policy fixed.

Frozen policies are evaluated using five matched test realizations per training seed and setting, all distinct from training. Outcomes are first averaged across the five test realizations within each training seed and then summarized across the five training seeds using two-sided 95\% Student $t$ confidence intervals. Percentage changes are paired within training seed. Checkpoint selection is based on training system reward subject to the joint completion criterion. Incomplete trips contribute travel time and exposure until timeout.

\section{Results}
\label{sec}

\subsection{Training Performance of Reinforcement Learning}

SARSA reached a relatively stable policy within the fixed 100-episode training budget across all five demand levels. During the early training stage, vehicle rewards increased rapidly, while micro-mobility rewards became progressively less negative. As exploration decayed, both reward components entered relatively stable regimes. Similar training patterns were observed across demand levels, indicating empirical stabilization within the predefined training budget. The selected policies achieved complete arrival for both vehicle and micro-mobility agents across the evaluated demand levels. MAPPO showed a similar training pattern, with the corresponding results reported in Appendix~\ref{app:comparator_training_curves}.

\begin{inlinefigure}
\centering
\includegraphics[width=0.95\linewidth]{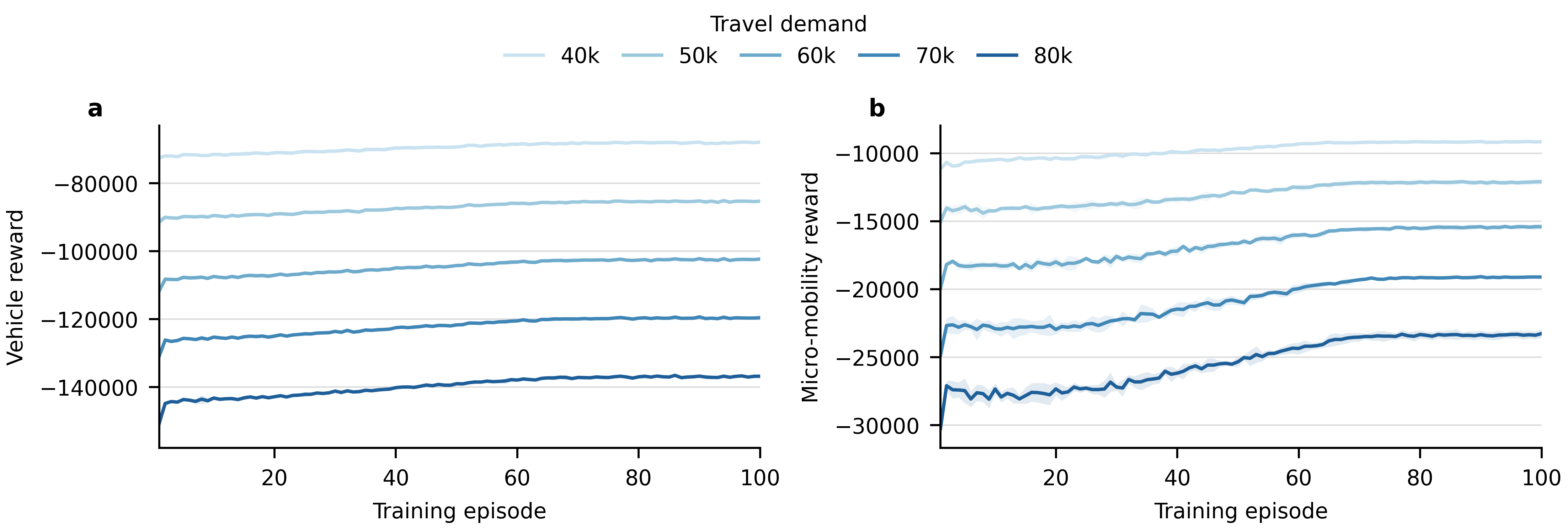}
\caption{SARSA training rewards on the 250 m grid. Lines and shaded bands represent five-seed means and 95\% confidence intervals, respectively.}
\label{fig:mfsarsa_training_curves}
\end{inlinefigure}

\subsection{Comparison with Baseline Routing Strategies}

SARSA achieved consistently strong joint performance across demand levels by substantially reducing both vehicle travel time and environmental exposure without materially increasing micro-mobility travel time. At $N=80{,}000$, relative to shortest-distance routing, SARSA reduced total vehicle travel time by 34.59\% and cumulative incremental exposure by 58.65\%, while total micro-mobility travel time remained nearly unchanged. At $N=40{,}000$, vehicle travel time and exposure decreased by 10.60\% and 57.79\%, respectively. Across demand levels, the exposure reduction remained stable at approximately 58--59\%, while the vehicle-time benefit increased with demand (Figure~\ref{fig:strategy_comparison}; Table~\ref{tab:routing_strategy_comparison}).

Dynamic exposure-aware routing also reduced vehicle travel time and exposure without a substantial micro-mobility time penalty. At $N=80{,}000$, it reduced vehicle travel time by 27.40\% and exposure by 39.15\%. SARSA further increased these reductions to 34.59\% and 58.65\%, respectively, indicating that the learned policy provided additional benefits beyond congestion-aware vehicle routing and dynamic exposure-aware micro-mobility routing.

The minimum-exposure baseline achieved the largest single reduction in exposure. At $N=80{,}000$, exposure decreased by 63.76\%, slightly more than the 58.65\% reduction achieved by SARSA. However, this was accompanied by a 46.35\% increase in total micro-mobility travel time. SARSA therefore retained most of the exposure benefit while avoiding this substantial time penalty, resulting in a better balance among vehicle efficiency, micro-mobility efficiency, and environmental exposure.

MAPPO also reduced vehicle travel time and exposure, but its improvements were generally smaller than those of SARSA. At $N=40{,}000$ and $50{,}000$, exposure decreased by 43.59\% and 39.45\%, respectively, while vehicle travel time decreased by 5.89\% and 10.43\%. At $N=80{,}000$, vehicle travel time and exposure decreased by 22.14\% and 22.64\%, respectively.

\begin{inlinefigure}
\centering
\includegraphics[width=0.95\textwidth]{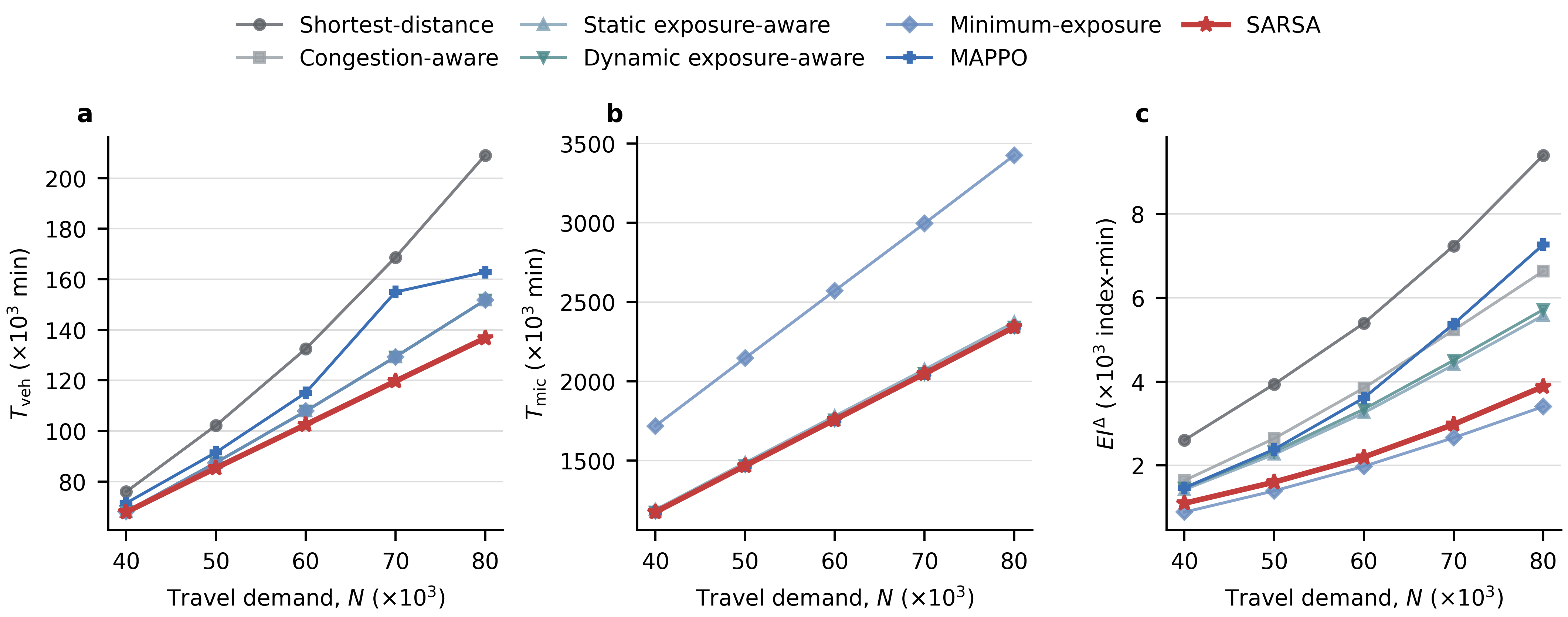}
\caption{Routing-strategy performance on the 250 m grid, including vehicle travel time, micro-mobility travel time, and environmental exposure. Points represent five-seed means, with five matched tests per seed.}
\label{fig:strategy_comparison}

\end{inlinefigure}

\begin{table}[pos=!htbp]\centering\scriptsize
\caption{Routing-strategy performance on the 250 m grid. Values are five-seed means [95\% confidence intervals], with five tests per seed. Time: $\times10^3$ min; exposure: $\times10^3$ index-min. Changes are relative to shortest-distance routing; totals include incomplete trips.}
\label{tab:routing_strategy_comparison}
\setlength{\tabcolsep}{1.6pt}\renewcommand{\arraystretch}{1.08}
\resizebox{\textwidth}{!}{%
\begin{tabular}{llrrrrrr}\toprule
$N$ & Method & \shortstack{$T_{\mathrm{veh}}$\\mean [95\% CI]} & \shortstack{$\Delta T_{\mathrm{veh}}$\\(\%)} & \shortstack{$T_{\mathrm{mic}}$\\mean [95\% CI]} & \shortstack{$\Delta T_{\mathrm{mic}}$\\(\%)} & \shortstack{$EI^{\Delta}$\\mean [95\% CI]} & \shortstack{$\Delta EI^{\Delta}$\\(\%)} \\ \midrule
40,000 & Shortest-distance & 76.0 [75.8, 76.2] & -- & 1173.4 [1173.4, 1173.4] & -- & 2.6 [2.6, 2.6] & -- \\
 & Congestion-aware & 68.1 [68.1, 68.1] & -10.33 & 1173.4 [1173.4, 1173.4] & +0.00 & 1.6 [1.6, 1.6] & -36.98 \\
 & Static exposure-aware & 68.1 [68.1, 68.1] & -10.32 & 1188.4 [1188.4, 1188.4] & +1.28 & 1.4 [1.4, 1.4] & -45.33 \\
 & Dynamic exposure-aware & 68.1 [68.1, 68.1] & -10.33 & 1173.4 [1173.4, 1173.4] & +0.00 & 1.5 [1.5, 1.5] & -43.77 \\
 & Minimum-exposure & 68.1 [68.1, 68.1] & -10.32 & 1717.0 [1717.0, 1717.0] & +46.34 & 0.9 [0.9, 0.9] & -65.92 \\
 & MAPPO & 71.5 [68.6, 74.4] & -5.89 & 1173.4 [1173.4, 1173.4] & +0.00 & 1.5 [1.4, 1.6] & -43.59 \\
\textbf{} & \textbf{SARSA} & \textbf{67.9 [67.9, 67.9]} & \textbf{-10.60} & \textbf{1173.4 [1173.4, 1173.4]} & \textbf{+0.00} & \textbf{1.1 [1.1, 1.1]} & \textbf{-57.79} \\
\midrule
50,000 & Shortest-distance & 102.2 [102.0, 102.5] & -- & 1465.9 [1465.9, 1465.9] & -- & 3.9 [3.9, 3.9] & -- \\
 & Congestion-aware & 87.5 [87.5, 87.5] & -14.40 & 1465.9 [1465.9, 1465.9] & +0.00 & 2.6 [2.6, 2.6] & -32.86 \\
 & Static exposure-aware & 87.5 [87.5, 87.5] & -14.40 & 1484.9 [1484.9, 1484.9] & +1.30 & 2.3 [2.3, 2.3] & -42.45 \\
 & Dynamic exposure-aware & 87.5 [87.5, 87.5] & -14.40 & 1465.9 [1465.9, 1465.9] & +0.00 & 2.3 [2.3, 2.3] & -40.73 \\
 & Minimum-exposure & 87.5 [87.5, 87.5] & -14.40 & 2145.5 [2145.5, 2145.5] & +46.36 & 1.4 [1.4, 1.4] & -64.57 \\
 & MAPPO & 91.5 [88.9, 94.2] & -10.43 & 1465.9 [1465.9, 1465.9] & +0.00 & 2.4 [2.2, 2.6] & -39.45 \\
\textbf{} & \textbf{SARSA} & \textbf{85.4 [85.2, 85.6]} & \textbf{-16.44} & \textbf{1465.9 [1465.9, 1465.9]} & \textbf{+0.00} & \textbf{1.6 [1.5, 1.7]} & \textbf{-59.28} \\
\midrule
60,000 & Shortest-distance & 132.5 [132.2, 132.8] & -- & 1757.2 [1757.2, 1757.2] & -- & 5.4 [5.4, 5.4] & -- \\
 & Congestion-aware & 107.9 [107.9, 107.9] & -18.58 & 1757.2 [1757.2, 1757.2] & +0.00 & 3.8 [3.8, 3.9] & -28.65 \\
 & Static exposure-aware & 107.9 [107.9, 107.9] & -18.58 & 1780.0 [1780.0, 1780.0] & +1.30 & 3.3 [3.2, 3.3] & -39.63 \\
 & Dynamic exposure-aware & 107.9 [107.9, 107.9] & -18.58 & 1757.2 [1757.2, 1757.2] & +0.00 & 3.3 [3.3, 3.3] & -38.05 \\
 & Minimum-exposure & 107.9 [107.8, 107.9] & -18.58 & 2571.8 [2571.8, 2571.8] & +46.35 & 2.0 [2.0, 2.0] & -63.28 \\
 & MAPPO & 115.1 [111.5, 118.7] & -13.12 & 1757.2 [1757.2, 1757.2] & +0.00 & 3.6 [3.1, 4.1] & -33.02 \\
\textbf{} & \textbf{SARSA} & \textbf{102.5 [102.5, 102.5]} & \textbf{-22.63} & \textbf{1757.2 [1757.2, 1757.2]} & \textbf{+0.00} & \textbf{2.2 [2.2, 2.3]} & \textbf{-59.18} \\
\midrule
70,000 & Shortest-distance & 168.7 [167.9, 169.5] & -- & 2047.8 [2047.8, 2047.8] & -- & 7.2 [7.2, 7.3] & -- \\
 & Congestion-aware & 129.3 [129.3, 129.4] & -23.32 & 2047.8 [2047.8, 2047.8] & +0.00 & 5.2 [5.2, 5.2] & -27.75 \\
 & Static exposure-aware & 129.3 [129.3, 129.4] & -23.32 & 2074.2 [2074.2, 2074.2] & +1.29 & 4.4 [4.4, 4.4] & -39.21 \\
 & Dynamic exposure-aware & 129.3 [129.3, 129.4] & -23.32 & 2047.8 [2047.8, 2047.8] & +0.00 & 4.5 [4.5, 4.5] & -37.70 \\
 & Minimum-exposure & 129.3 [129.3, 129.4] & -23.32 & 2996.5 [2996.5, 2996.5] & +46.33 & 2.7 [2.7, 2.7] & -63.21 \\
 & MAPPO & 155.0 [86.8, 223.1] & -8.10 & 2047.8 [2047.8, 2047.8] & +0.00 & 5.4 [4.3, 6.4] & -25.75 \\
\textbf{} & \textbf{SARSA} & \textbf{119.7 [119.6, 119.8]} & \textbf{-29.02} & \textbf{2047.8 [2047.8, 2047.8]} & \textbf{+0.00} & \textbf{3.0 [3.0, 3.0]} & \textbf{-58.82} \\
\midrule
80,000 & Shortest-distance & 209.1 [207.9, 210.2] & -- & 2341.8 [2341.8, 2341.8] & -- & 9.4 [9.4, 9.4] & -- \\
 & Congestion-aware & 151.8 [151.7, 151.8] & -27.40 & 2341.8 [2341.8, 2341.8] & +0.00 & 6.6 [6.6, 6.6] & -29.39 \\
 & Static exposure-aware & 151.8 [151.7, 151.8] & -27.41 & 2372.0 [2372.0, 2372.0] & +1.29 & 5.6 [5.6, 5.6] & -40.64 \\
 & Dynamic exposure-aware & 151.8 [151.7, 151.8] & -27.40 & 2341.8 [2341.8, 2341.8] & +0.00 & 5.7 [5.7, 5.7] & -39.15 \\
 & Minimum-exposure & 151.8 [151.7, 151.8] & -27.41 & 3427.1 [3427.1, 3427.1] & +46.35 & 3.4 [3.4, 3.4] & -63.76 \\
 & MAPPO & 162.8 [150.3, 175.3] & -22.14 & 2341.8 [2341.8, 2341.8] & +0.00 & 7.3 [6.1, 8.5] & -22.64 \\
\textbf{} & \textbf{SARSA} & \textbf{136.7 [136.7, 136.8]} & \textbf{-34.59} & \textbf{2341.8 [2341.8, 2341.8]} & \textbf{+0.00} & \textbf{3.9 [3.8, 4.0]} & \textbf{-58.65} \\
\bottomrule\end{tabular}
}
\end{table}

\subsection{Spatial Flow and Exposure Patterns}

SARSA jointly redistributed vehicle and micro-mobility routes, reducing both traffic concentration and environmental exposure hotspots. Under shortest-distance routing, both modes were concentrated along a limited number of direct corridors, causing high traffic loads and exposure hotspots to overlap on several links. Dynamic exposure-aware routing diverted micro-mobility users away from some high-exposure links but did not explicitly coordinate this redistribution with vehicle congestion (Figure~\ref{fig:spatial_comparison_primary}).

In contrast, SARSA redistributed part of the vehicle flow away from highly concentrated corridors while guiding micro-mobility users away from locations with high congestion-induced exposure. The resulting exposure reduction therefore came from both congestion mitigation and low-exposure micro-mobility routing, rather than from substantial micro-mobility detours alone. This spatial pattern is consistent with the aggregate reductions in vehicle travel time and exposure while micro-mobility travel time remained nearly unchanged (Figure~\ref{fig:spatial_comparison_primary}). Spatial results for the other four strategies are reported in Appendix~\ref{app:additional_spatial}.

\begin{inlinefigure}
\centering
\includegraphics[
width=0.98\textwidth,
height=0.60\textheight,
keepaspectratio
]{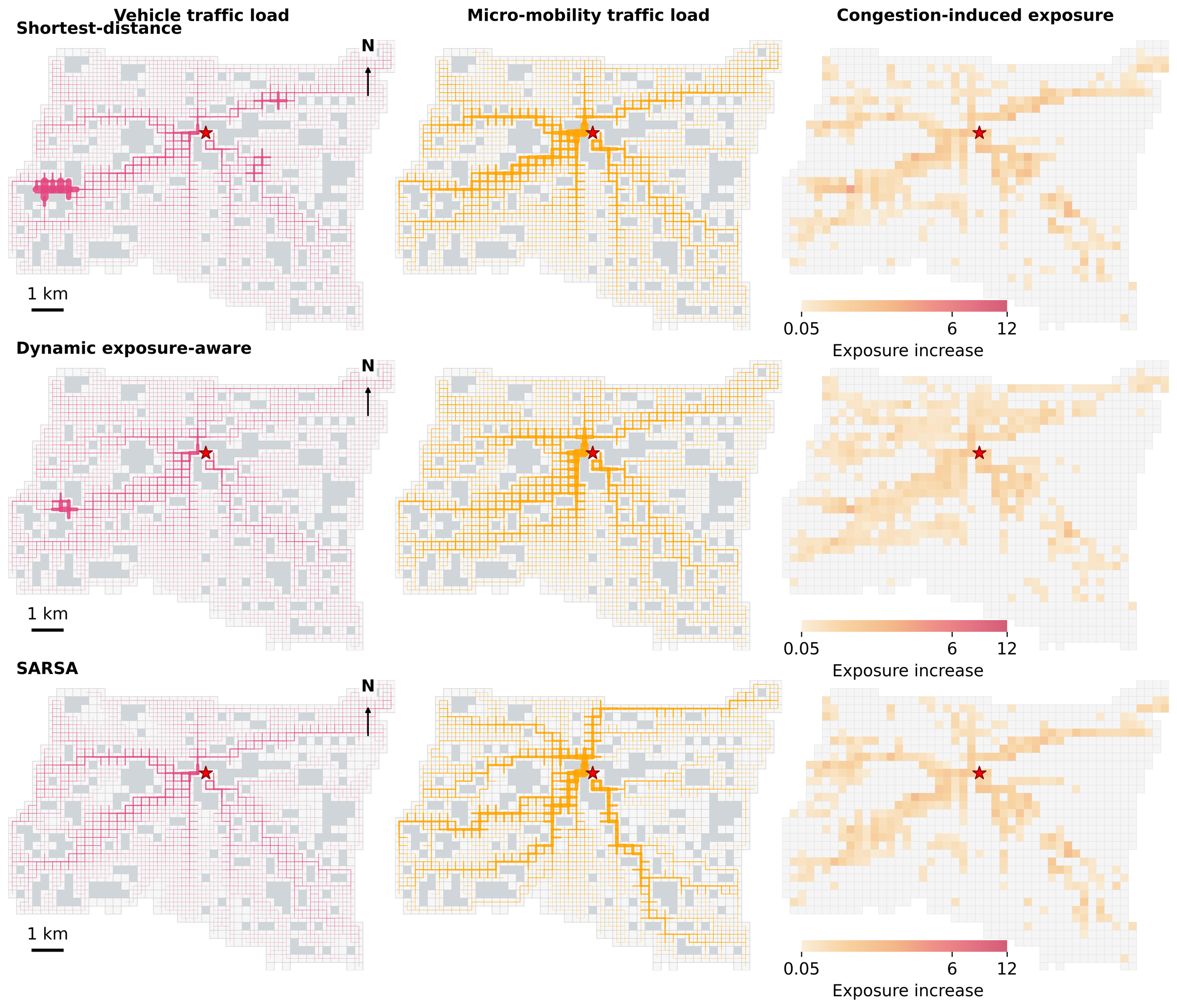}
\caption{Spatial patterns from the first frozen-policy test at $N=80{,}000$ for seed 42 under shortest-distance routing, dynamic exposure-aware routing, and SARSA. Columns show vehicle traffic load, micro-mobility traffic load, and congestion-induced exposure.}
\label{fig:spatial_comparison_primary}
\end{inlinefigure}

\subsection{Effect of Environmental Exposure Constraints}

Congestion avoidance by vehicles already reduced part of the environmental exposure by alleviating local traffic concentration, but additional low-exposure routing for micro-mobility users produced substantial further benefits. Under the setting without an explicit exposure term ($\eta=0$), vehicles still adjusted their routes according to congestion conditions. The resulting redistribution of traffic therefore reduced part of the traffic-related exposure even without an explicit micro-mobility exposure objective.

However, congestion mitigation alone did not achieve the full exposure reduction. When environmental exposure was further incorporated into micro-mobility routing, users could actively avoid remaining high-exposure links. Relative to $\eta=0$, cumulative incremental exposure decreased by a further 38.77\% at $\eta=1$ and by 39.73\% under the main setting of $\eta=3$, while total micro-mobility travel time remained nearly unchanged (Table~\ref{tab:sarsa_exposure_weight}). This indicates that low-exposure micro-mobility routing remained effective even after part of the exposure had already been reduced through vehicle congestion mitigation.

Further increases in the exposure weight produced only small marginal gains. Relative to $\eta=0$, exposure decreased by 39.77\% at $\eta=5$ and 40.48\% at $\eta=7$. The main benefit therefore came from explicitly incorporating exposure into micro-mobility routing rather than continuously increasing its weight. The main setting of $\eta=3$ captured most of the additional benefit associated with exposure-aware routing.

\begin{table}[pos=!htbp]\centering\small
\caption{SARSA exposure-weight sensitivity at $N=80{,}000$ on the 250 m grid. Values are five-seed means [95\% confidence intervals], with five tests per seed. Time: $\times10^3$ min; exposure: $\times10^3$ index-min. Changes are relative to $\eta=0$.}
\label{tab:sarsa_exposure_weight}
\setlength{\tabcolsep}{2.0pt}\renewcommand{\arraystretch}{1.20}
\begin{tabular*}{\textwidth}{@{\extracolsep{\fill}}rrrrr@{}}\toprule
$\eta$ & \shortstack{$T_{\mathrm{mic}}$\\mean [95\% CI]} & \shortstack{$\Delta T_{\mathrm{mic}}$\\mean (\%)} & \shortstack{$EI^{\Delta}$\\mean [95\% CI]} & \shortstack{$\Delta EI^{\Delta}$\\mean (\%)} \\ \midrule
0.0 & 2341.8 [2341.8, 2341.8] & -- & 6.5 [6.2, 6.7] & -- \\
1.0 & 2341.8 [2341.8, 2341.8] & +0.00 & 3.9 [3.9, 4.0] & -38.77 \\
3.0 & 2341.8 [2341.8, 2341.8] & +0.00 & 3.9 [3.8, 4.0] & -39.73 \\
5.0 & 2341.8 [2341.8, 2341.8] & +0.00 & 3.9 [3.8, 4.0] & -39.77 \\
7.0 & 2341.8 [2341.8, 2341.8] & +0.00 & 3.8 [3.8, 3.9] & -40.48 \\
\bottomrule\end{tabular*}
\end{table}

\subsection{Sensitivity to Route-Guidance Obedience}

The benefits of SARSA did not require all travelers to fully follow the learned route guidance. In the obedience sensitivity experiment, vehicles that did not adopt SARSA followed congestion-aware routing, while non-adopting micro-mobility users followed shortest-distance routing. As obedience increased, a larger share of travelers switched from these default strategies to SARSA guidance. The experiment therefore measures the additional benefit of SARSA under different levels of route-guidance adoption.

Partial adoption already preserved substantial traffic and environmental benefits, while higher obedience produced further exposure reduction (Table~\ref{tab:obedience_sensitivity}). Even at 0\% obedience, mean exposure was 29.49\% lower than under the shortest-distance baseline because vehicles still followed congestion-aware routing. As obedience increased, exposure reduction strengthened progressively and reached 58.65\% at full obedience. This indicates that coordinated vehicle routing and low-exposure micro-mobility guidance provide additional exposure benefits beyond congestion mitigation alone.

Vehicle-efficiency gains saturated earlier than exposure reductions. From 0\% to 100\% obedience, total vehicle travel time decreased by a further 9.94\%, with most of this benefit already achieved at 75\% obedience. In contrast, environmental exposure continued to decline as obedience increased. Total micro-mobility travel time remained nearly unchanged across all obedience levels because candidate actions were constrained to maintain progress toward the destination. Overall, the framework retains substantial benefits under partial adoption and does not rely on the idealized assumption of complete compliance.

\begin{table}[pos=!htbp]\centering\small
\caption{SARSA obedience sensitivity at $N=80{,}000$ on the 250 m grid. Values are five-seed means [95\% confidence intervals], with five tests per seed. Time: $\times10^3$ min; exposure: $\times10^3$ index-min. Exposure changes are relative to shortest-distance routing; realized rates denote vehicle/micro-mobility obedience.}
\label{tab:obedience_sensitivity}
\setlength{\tabcolsep}{1.6pt}\renewcommand{\arraystretch}{1.20}
\begin{tabular*}{\textwidth}{@{\extracolsep{\fill}}rrrrrr@{}}\toprule
\shortstack{Target\\obedience (\%)} & \shortstack{Realized obedience\\vehicle/micro (\%)} & \shortstack{$T_{\mathrm{veh}}$\\mean [95\% CI]} & \shortstack{$T_{\mathrm{mic}}$\\mean [95\% CI]} & \shortstack{$EI^{\Delta}$\\mean [95\% CI]} & \shortstack{$\Delta EI^{\Delta}$\\mean (\%)} \\ \midrule
0 & 0.00/0.00 & 151.8 [151.8, 151.8] & 2341.8 [2341.8, 2341.8] & 6.6 [6.6, 6.6] & -29.49 \\
25 & 25.08/25.02 & 143.2 [142.5, 143.8] & 2341.8 [2341.8, 2341.8] & 6.4 [6.4, 6.5] & -31.47 \\
50 & 50.10/49.98 & 138.3 [137.5, 139.0] & 2341.8 [2341.8, 2341.8] & 6.0 [5.9, 6.1] & -36.32 \\
75 & 75.21/74.98 & 136.6 [136.4, 136.8] & 2341.8 [2341.8, 2341.8] & 5.3 [5.2, 5.3] & -43.79 \\
100 & 100.00/100.00 & 136.7 [136.7, 136.8] & 2341.8 [2341.8, 2341.8] & 3.9 [3.8, 4.0] & -58.65 \\
\bottomrule\end{tabular*}
\end{table}
\section{Discussion}

Mega-event travel represents a short-term, high-demand setting in which large numbers of travelers make route choices within a narrow time window and compete for limited road capacity. Under these conditions, route choices are strongly interdependent. When many travelers select similar routes, local traffic conditions can change rapidly, which then affects the travel costs faced by other travelers. Rule-based routing can respond to current conditions but has limited ability to account for the cumulative system effects generated by repeated interactions among many agents \citep{ke2025realtime}. Multi-agent reinforcement learning is therefore particularly relevant in mega-event scenarios because it can learn from the evolving consequences of collective route choices \citep{shou2022multi,wang2025scalable}. This becomes increasingly important as travel demand rises and the gap between individual route decisions and system-level outcomes becomes more pronounced. The compact mean-field representation used by SARSA further provides a scalable way to summarize nearby traffic pressure without explicitly representing every other agent.

Congestion mitigation for vehicles and exposure reduction for micro-mobility users operate through different but complementary mechanisms. Redistributing vehicle routes can reduce local traffic concentration and simultaneously lower part of the traffic-related exposure, indicating an environmental co-benefit of improved traffic efficiency. However, local high-exposure links may remain even after congestion is reduced, and the exposure experienced by micro-mobility users still depends on the specific roads traveled, exposure duration, and local traffic conditions. The exposure-weight experiment further shows that explicitly considering micro-mobility exposure can provide additional reductions after congestion has already been improved. Lower system-wide congestion therefore does not automatically guarantee lower environmental risk along the routes experienced by micro-mobility users, and health-oriented transport optimization still needs to directly account for individual exposure pathways \citep{hertel2008human,hatzopoulou2013web,rafiepourgatabi2021route,wu2022pm25}. Low-exposure routing should also remain subject to efficiency constraints. By restricting candidate actions to keep travelers moving toward their destinations, the proposed framework controls detours while optimizing environmental exposure and avoids obtaining lower exposure through excessively long routes. Vehicle users are mainly affected by congestion and travel delay, whereas micro-mobility users are more directly exposed to traffic-related air pollution and noise. Differentiated objectives within a shared road network can therefore better represent the actual costs and risks faced by different travel modes.

The practical effects of route guidance also depend on traveler adoption. Participants in mega events differ in route preferences, information sources, and willingness to follow navigation guidance, making complete compliance with a common routing strategy unrealistic. The obedience analysis shows that substantial traffic and environmental benefits can still be retained when only part of the traveling population adopts the guidance, while higher adoption can further reduce environmental exposure. When some travelers change their routes, they also reshape network traffic loads and local exposure conditions, which can affect the travel environment experienced by other users. This network-level indirect effect suggests that behavioral intervention applied to only a subset of travelers may still generate broader system benefits \citep{xu2017collective}. From a planning perspective, the framework can support preventive traffic management for mega events. Simulations conducted before an event can identify key corridors that are likely to develop traffic concentrations and exposure hotspots under high demand, supporting temporary traffic organization, vehicle diversion, and micro-mobility route guidance \citep{villiers2019traffic,xing2025empirical,giuliano1990staggered}. Vehicle management can focus on relieving congestion along high-demand corridors, while cycling and walking guidance can further account for pollution and noise risks. Mega-event traffic management can therefore extend beyond a single efficiency-oriented objective toward differentiated management that also considers the environmental health needs of different travel modes.

This study also provides several directions for further extension. The current framework dynamically represents traffic-state changes, PM$_{2.5}$ and noise exposure, and cumulative exposure along travel routes, while some finer-scale environmental processes, such as complex meteorological conditions, street-canyon effects, acoustic barriers, and localized propagation, are not further resolved. Accordingly, the environmental module is primarily intended to support relative comparisons among routing strategies under matched background conditions, with a consistent representation of traffic-related environmental increments across scenarios. Future traffic simulations could also incorporate more detailed intersection control, heterogeneous route preferences, and observed event travel trajectories to improve adaptability across scenarios and traveler groups. Integrating real-time environmental monitoring and trajectory data during mega events could further support calibration of traffic--environment relationships and extend the framework toward practical traffic-management applications.

\section{Conclusion}

This study develops an exposure-aware multi-agent routing framework for short-term, high-demand travel scenarios, using mega-event travel as a representative case. The framework integrates dynamic traffic conditions and environmental exposure into differentiated route decisions for vehicle and micro-mobility agents.

The results show that the framework can substantially reduce environmental exposure while improving vehicle travel efficiency and maintaining micro-mobility travel efficiency. Across demand levels, cumulative environmental exposure decreased consistently by approximately 58--59\%, while the improvement in vehicle travel time became more pronounced as demand increased. Under the high-demand scenario, SARSA reduced total vehicle travel time by 34.59\% and cumulative environmental exposure for micro-mobility users by 58.65\% relative to shortest-distance routing, with little change in micro-mobility travel time. Exposure reduction resulted from two complementary mechanisms. Vehicle-side congestion mitigation reduced part of the traffic-related environmental burden, while low-exposure routing further guided micro-mobility users away from remaining high-exposure locations. This indicates that improving transportation efficiency can provide environmental co-benefits, but cannot fully substitute for explicitly incorporating environmental exposure into micro-mobility route decisions.

The framework extends conventional efficiency-oriented route optimization toward coordinated management of transportation efficiency and environmental health. It provides a quantitative basis for traffic organization and low-exposure route guidance in mega events and other short-term high-demand scenarios. Future work can incorporate observed event-specific travel demand and trajectories, more detailed road and intersection dynamics, and real-time environmental sensing to improve the framework's applicability to more complex urban networks.

\section*{Acknowledgements}

This work was supported by the Guangdong Provincial Project 
(Grant No. 2024QN11G197) and the National Natural Science Foundation 
of China (Grant Nos. T2600668 and 42301251).


\FloatBarrier
\appendix
\section*{Appendix}

\section{Study Area}
\label{app:study_area}

Figure~\ref{fig:study_area_appendix} shows the Olympic Sports Center and its surrounding road network used in the simulation.

\begin{inlinefigure}
\centering
\includegraphics[
    width=0.65\linewidth,
    height=0.55\textheight,
    keepaspectratio
]{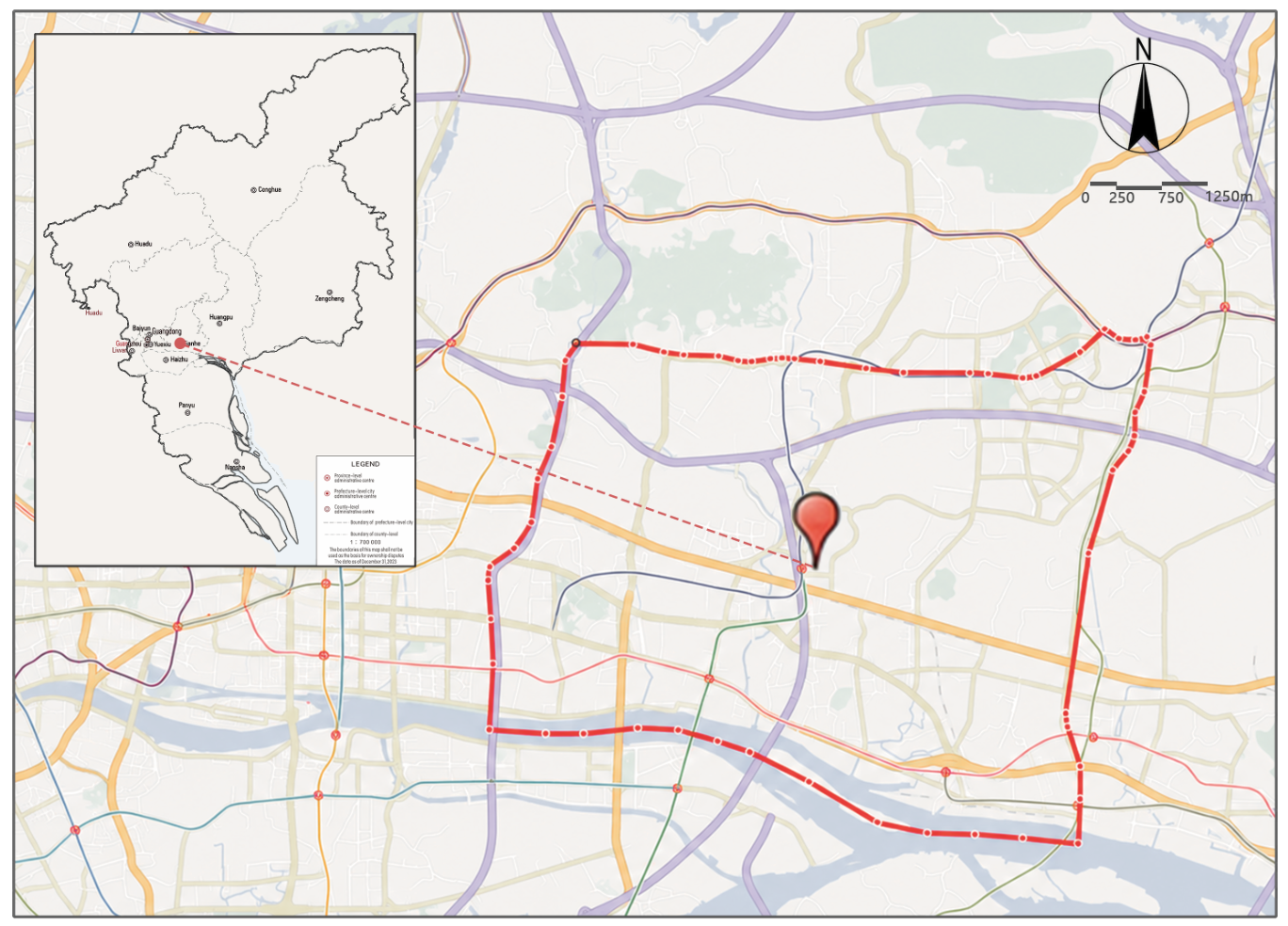}
\caption{Study area.}
\label{fig:study_area_appendix}
\end{inlinefigure}

\section{Training Curves of the Neural Comparator}
\label{app:comparator_training_curves}

Figure~\ref{fig:mappo_training_curves_appendix} shows the complete 100-episode reward trajectories for MAPPO on the 250 m grid. It uses the same operational time--exposure objectives, demand levels, seeds, uncertainty definition, and unsmoothed episode scale as the SARSA curves in Figure~\ref{fig:mfsarsa_training_curves}, with the algorithm-specific auxiliary shaping described in Section~3.

\begin{inlinefigure}
\centering
\includegraphics[width=0.95\linewidth]{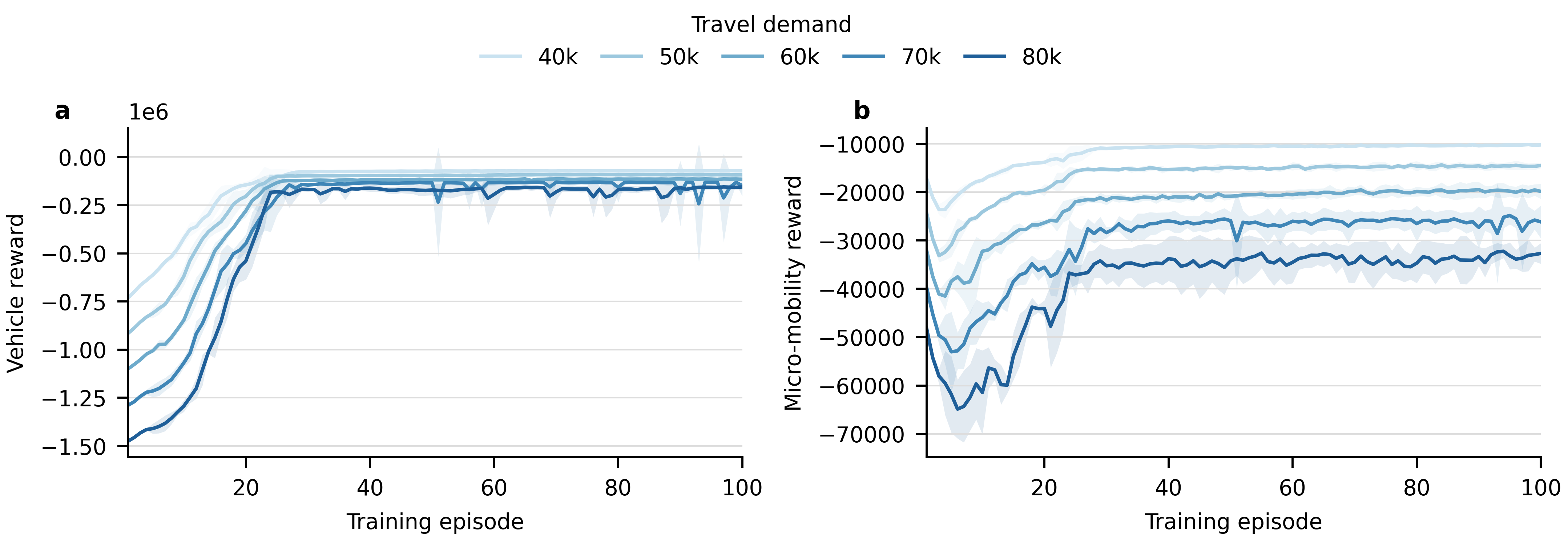}
\caption{MAPPO training rewards on the 250 m grid. Lines and shaded bands show five-seed means and 95\% confidence intervals.}
\label{fig:mappo_training_curves_appendix}
\end{inlinefigure}

\section{Reward Parameters}
\label{app:reward_parameters}

Table~\ref{tab:reward_parameters} lists the reward coefficients used in the reported SARSA experiments. Within each experimental setting, these coefficients were identical across all five training seeds. The 250 m SARSA settings were shared by the main and exposure-weight experiments; only $\eta$ varied in the latter. The main experiment used $\eta=3$, whereas the exposure-weight experiment used $\eta\in\{0,1,3,5,7\}$ at $N=80{,}000$.

The multiscale SARSA experiments used $\eta=3$ and the grid-specific vehicle coefficients listed in Table~\ref{tab:reward_parameters}. Micro-mobility time, progress, and arrival coefficients, as well as the timeout penalty, were unchanged across these settings. Time penalties are applied per minute; progress is measured in shortest-path grid steps, and the milestone term is based on normalized improvements in the best distance reached.

At 150 m, the milestone coefficient was set to zero, and an additional progress reward was defined using $\Phi(g)=-\beta d(g)$, with $\beta=500/36$. The vehicle reward includes the potential difference $\Phi(g_{k+1})-\Phi(g_k)$ between consecutive states. Ordinary transitions therefore receive $\beta\Delta d_{a,k}$, while timeout additionally returns $\beta d(g_{\mathrm{timeout}})$ to maintain zero terminal potential. The existing timeout penalty remains unchanged. The original Q-value initialization was retained, and the initial learning rate remained 0.06.

\begin{table}[pos=!htbp]
\centering\small
\caption{SARSA reward coefficients.}
\label{tab:reward_parameters}
\setlength{\tabcolsep}{4pt}
\renewcommand{\arraystretch}{1.15}
\begin{tabular}{lcccc}
\toprule
 & 250 m & \multicolumn{3}{c}{Multiscale} \\
\cmidrule(lr){2-2}\cmidrule(lr){3-5}
Coefficient & Main / exposure-weight & 100 m & 150 m & 200 m \\
\midrule
Vehicle time $w_{\mathrm{veh}}$ & 1 & 5 & 5 & 5 \\
Micro-mobility time $w_{\mathrm{mic}}$ & 0.005 & 0.005 & 0.005 & 0.005 \\
Exposure $\eta$ & $3^{\mathrm{a}}$ & 3 & 3 & 3 \\
Vehicle progress $w^{\mathrm{veh}}_p$ & 0.1 & 2.5 & $10/3$ & 1.25 \\
Additional vehicle progress $\beta$ & 0 & 0 & $500/36$ & 0 \\
Micro-mobility progress $w^{\mathrm{mic}}_p$ & 0.02 & 0.02 & 0.02 & 0.02 \\
Milestone $R_{\mathrm{ms}}$ & 0 & 300 & 0 & 300 \\
Non-progress $R_{\mathrm{np}}$ & 0 & 7.5 & 10 & 3.75 \\
Vehicle arrival $R^{\mathrm{veh}}_{\mathrm{goal}}$ & 200 & 520 & $1000/3$ & 250 \\
Micro-mobility arrival $R^{\mathrm{mic}}_{\mathrm{goal}}$ & 200 & 200 & 200 & 200 \\
Timeout $R_{\mathrm{timeout}}$ & 200 & 200 & 200 & 200 \\
\bottomrule
\end{tabular}

\smallskip
\begin{minipage}{0.95\linewidth}\small
$^{\mathrm{a}}$ The exposure-weight experiment varied only $\eta$ over $\{0,1,3,5,7\}$; all other coefficients matched the 250 m MF-SARSA column.
\end{minipage}
\end{table}

\section{Performance across Grid Scales}
\label{app:grid_resolution}

To examine the performance of the framework across different spatial resolutions, the high-demand scenario with $N=80{,}000$ was repeated using 100, 150, and 200 m grids. Changes in grid resolution affect spatial representation, road capacity, agent movement, and the scale of reward signals. Therefore, selected reward parameters were adjusted for each resolution to maintain stable learning and consistent reward interpretation across grid scales. The corresponding parameter settings are reported in Appendix~\ref{app:reward_parameters}. These experiments evaluate whether the framework can be adapted to different spatial resolutions while preserving its overall performance pattern.

Results show that, relative to the shortest-distance baseline, SARSA reduced vehicle travel time by 28.65\%, 22.66\%, and 46.31\% on the 100, 150, and 200 m grids, respectively, while cumulative incremental exposure decreased by 63.82\%, 61.65\%, and 53.70\%. Micro-mobility travel time remained essentially unchanged across all three resolutions. The direction of improvement in traffic efficiency and environmental exposure was consistent across grid scales, although the magnitude of the effects varied with spatial resolution. Overall, the results show that the framework can be applied across multiple spatial resolutions with resolution-specific parameter adjustment while maintaining consistent improvements in vehicle travel efficiency and environmental exposure.

\begin{table}[pos=!htbp]\centering\scriptsize
\caption{Grid-resolution sensitivity at $N=80{,}000$. Values are five-seed means [95\% confidence intervals], with five matched tests per seed. Time: $\times10^3$ min; exposure: $\times10^3$ index-min. Changes are relative to shortest-distance routing; totals include incomplete trips.}
\label{tab:multiscale}
\setlength{\tabcolsep}{1.6pt}\renewcommand{\arraystretch}{1.08}
\resizebox{\textwidth}{!}{%
\begin{tabular}{llrrrrrr}\toprule
Grid (m) & Method & \shortstack{$T_{\mathrm{veh}}$\\mean [95\% CI]} & \shortstack{$\Delta T_{\mathrm{veh}}$\\(\%)} & \shortstack{$T_{\mathrm{mic}}$\\mean [95\% CI]} & \shortstack{$\Delta T_{\mathrm{mic}}$\\(\%)} & \shortstack{$EI^{\Delta}$\\mean [95\% CI]} & \shortstack{$\Delta EI^{\Delta}$\\(\%)} \\ \midrule
100 & Shortest-distance & 470.2 [468.7, 471.7] & -- & 2762.7 [2761.7, 2763.7] & -- & 34.2 [34.2, 34.3] & -- \\
 & Congestion-aware & 402.6 [401.8, 403.3] & -14.38 & 2762.7 [2761.7, 2763.7] & +0.00 & 30.8 [30.6, 31.0] & -10.18 \\
 & Static exposure-aware & 404.0 [402.3, 405.8] & -14.07 & 2782.3 [2781.2, 2783.4] & +0.71 & 19.6 [19.5, 19.7] & -42.77 \\
 & Dynamic exposure-aware & 402.6 [401.8, 403.3] & -14.38 & 2762.9 [2761.9, 2763.9] & +0.00 & 22.3 [22.1, 22.5] & -34.83 \\
 & Minimum-exposure & 403.9 [402.3, 405.6] & -14.09 & 4105.4 [4102.2, 4108.6] & +48.60 & 8.7 [8.7, 8.7] & -74.67 \\
\textbf{} & \textbf{SARSA} & \textbf{335.5 [321.9, 349.1]} & \textbf{-28.65} & \textbf{2762.7 [2761.7, 2763.7]} & \textbf{+0.00} & \textbf{12.4 [10.5, 14.3]} & \textbf{-63.82} \\
\midrule
150 & Shortest-distance & 279.5 [279.2, 279.7] & -- & 2523.2 [2523.2, 2523.2] & -- & 19.7 [19.6, 19.7] & -- \\
 & Congestion-aware & 246.1 [245.9, 246.3] & -11.95 & 2523.2 [2523.2, 2523.2] & +0.00 & 15.0 [15.0, 15.1] & -23.73 \\
 & Static exposure-aware & 246.1 [245.8, 246.4] & -11.94 & 2550.1 [2550.1, 2550.1] & +1.07 & 13.9 [13.9, 14.0] & -29.31 \\
 & Dynamic exposure-aware & 246.1 [245.9, 246.3] & -11.95 & 2523.2 [2523.2, 2523.2] & +0.00 & 12.9 [12.9, 12.9] & -34.61 \\
 & Minimum-exposure & 246.1 [246.0, 246.3] & -11.94 & 4097.6 [4097.6, 4097.6] & +62.40 & 11.0 [11.0, 11.1] & -43.90 \\
\textbf{} & \textbf{SARSA} & \textbf{216.1 [208.5, 223.8]} & \textbf{-22.66} & \textbf{2523.2 [2523.2, 2523.2]} & \textbf{+0.00} & \textbf{7.6 [6.9, 8.2]} & \textbf{-61.65} \\
\midrule
200 & Shortest-distance & 339.1 [338.2, 340.0] & -- & 2361.8 [2361.8, 2361.8] & -- & 15.2 [15.2, 15.3] & -- \\
 & Congestion-aware & 189.3 [189.2, 189.5] & -44.16 & 2361.8 [2361.8, 2361.8] & +0.00 & 9.3 [9.3, 9.3] & -38.90 \\
 & Static exposure-aware & 189.3 [189.2, 189.5] & -44.16 & 2365.8 [2365.8, 2365.8] & +0.17 & 8.6 [8.5, 8.6] & -43.81 \\
 & Dynamic exposure-aware & 189.3 [189.2, 189.5] & -44.16 & 2361.8 [2361.8, 2361.8] & +0.00 & 8.2 [8.2, 8.2] & -46.24 \\
 & Minimum-exposure & 189.3 [189.2, 189.4] & -44.16 & 3663.5 [3663.5, 3663.5] & +55.12 & 4.3 [4.3, 4.3] & -71.70 \\
\textbf{} & \textbf{SARSA} & \textbf{182.1 [178.9, 185.2]} & \textbf{-46.31} & \textbf{2361.8 [2361.8, 2361.8]} & \textbf{+0.00} & \textbf{7.1 [6.2, 7.9]} & \textbf{-53.70} \\
\bottomrule\end{tabular}
}
\end{table}
\section{Spatial Diagnostics for Additional Routing Methods}
\label{app:additional_spatial}

Figure~\ref{fig:spatial_comparison_additional} provides the remaining four representative seed-42 spatial diagnostics under the high-demand setting: congestion-aware routing, static exposure-aware routing, minimum-exposure routing, and MAPPO.

\begin{inlinefigure}
\centering
\includegraphics[
    width=\linewidth,
    height=\textheight,
    keepaspectratio
]{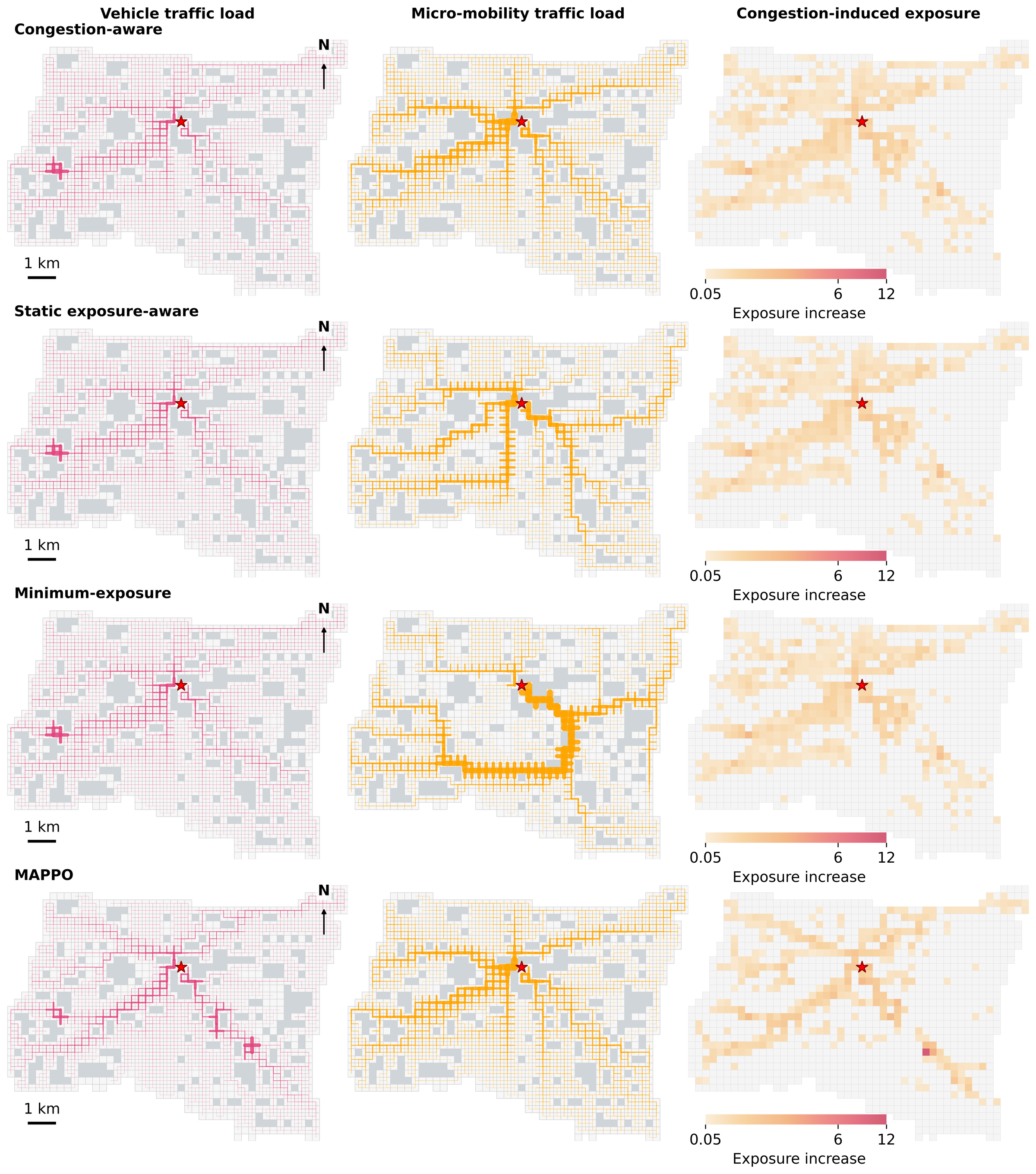}
\caption{Test spatial patterns for the four additional routing strategies, using the same columns, conditions, and scales as Figure~\ref{fig:spatial_comparison_primary}.}
\label{fig:spatial_comparison_additional}
\end{inlinefigure}
\FloatBarrier

During the preparation of this work the authors used Chatgpt in order to polish the grammar. After using this tool, the authors reviewed and edited the content as needed and take full responsibility for the content of the publication.
\bibliographystyle{cas-model2-names}
\bibliography{cas-refs}


\end{document}